\documentclass[prodmode]{acmsmall}          
\usepackage{graphicx}
\usepackage{amssymb}
\title{Reuse, Temporal Dynamics, Interest Sharing, and Collaboration in Social Tagging Systems
}

\author{ELIZEU SANTOS-NETO
\affil{University of British Columbia}
DAVID CONDON
\affil{University of South Florida}
NAZARENO ANDRADE
\affil{Universidade Federal de Campina Grande}
ADRIANA IAMNITCHI
\affil{University of South Florida}
MATEI RIPEANU
\affil{University of British Columbia}}

\begin{abstract}
User-generated content is shaping the dynamics of the World Wide Web. Indeed, an increasingly large number of systems provide mechanisms to support the growing demand for content creation, sharing, and management. Tagging systems are a particular class of these systems where users share and collaboratively annotate content such as photos and URLs. This collaborative behavior and the pool of user-generated metadata create opportunities to improve existing systems and to design new mechanisms. However, to realize this potential, it is necessary to first understand the usage characteristics of current systems.

This work addresses this issue characterizing three tagging systems ({\em CiteULike}, {\em Connotea} and {\em del.icio.us}) while focusing on three aspects: {\em i}) the patterns of information (tags and items) production; {\em ii}) the temporal dynamics of users' tag vocabularies; and, {\em iii}) the social aspects of tagging systems. The analysis of the patterns of information production shows that users publish new content more often than they annotate already existing content in the system. The opposite, however, occurs for tags; the level of tag reuse is much higher. This observation provides evidence that tags are indeed used for categorization. The relative difference between the rate of item publication and tag reuse suggests that tags are potentially useful as an additional source of information for item recommendation techniques.

The study of the temporal dynamics of user vocabularies shows that the growth rate of tag vocabularies across the user population over time decreases at early ages, stabilizes, and returns to increase for older users. Moreover, a closer look into the change of vocabulary contents over time shows that despite the fact that tag vocabularies are slowly growing in size with user age, the relative frequency in which each tag is used converges relatively quickly in a user’s lifetime. Mechanisms that rely on tag-based user similarity offer opportunities to harness the above observation by attempting to strike a balance between the accuracy of vocabulary similarity estimates, the data volume required for estimation, and the freshness of the data used.

Finally, the characterization of social aspects of tagging unveils the relationship between the implicit user ties, as inferred from the similarity between users' activity, and their explicit social ties, as represented by co-membership in discussion groups or semantic similarity between tag vocabularies. In particular, the results show that mechanisms that aim to harness the social ties between users can exploit the fact that implicit social ties complement their explicit counterparts with finer-grained strength information.
\end{abstract}

\begin{document}
\maketitle

\section{Introduction}

Tagging systems~\cite{Mathes2004,Hammond2005,Marlow2006,Macgregor2006,Farooq2007} are a ubiquitous manifestation of online peer-production of information~\cite{Benkler2006}, a production mode commonplace in today's World Wide Web~\cite{Ramakrishnan2007}. The annotation feature, often referred to as simply tagging, has been originally designed to support personal content management. However, as this feature exposes user preferences and their temporal dynamics, similarities between users, and the aggregated characteristics of the user population, annotations have been recognized for their potential to support a wider range of mechanisms such as social search~\cite{Yahia2008}, recommendation~\cite{Sigurbjornsson2008}, and search optimization~\cite{Yanbe2007,Heymann2008,Huang2008}.

Moreover, tagging is increasingly important in online social systems and, more recently, motivates new initiatives such as {\em OpenAnnotation}~\footnote{http://openannotation.org} that aims to enable users to annotate content on the web without depending on specific systems. Therefore, understanding social tagging through characterization and modeling of usage patterns is important, as understanding the current systems can better inform the design of future annotation platforms such as {\em Hypothes.is}~\footnote{http://hypothes.is}. Finally, characterizing social tagging systems can both unveil new opportunities and improve existing mechanisms.

This work extends our previous study~\cite{Santos-Neto2009} and addresses this need for characterization by investigating unexplored aspects of social tagging behaviour as well as complementing previous characterization studies (presented in Section~\ref{sec:related_work}). In particular, it focuses on three major aspects of the tagging activity that have attracted relatively little attention in the past: {\em i}) the dynamics of tag and items produced via collaborative annotation; {\em ii}) the temporal dynamics of users' tag vocabularies; and, {\em iii}) the characteristics of the social ties between users in these systems. A marked difference from this work to previously similar characterization studies is that, this study takes one step further by offering observations across multiple social tagging systems, which allows for a richer analysis of tagging behaviour.

To study the productions of tags and items, Section~\ref{sec:reuse} concentrates on two metrics: {\em i}) item re-tagging, a measure of the degree to which items are repeatedly tagged; and {\em ii}) tag reuse, a measure of the degree to which users reuse a tag to perform new annotations.

The analysis of the evolution of the users tag vocabularies (i.e., the set of tags a user assigns to her items) in Section~\ref{sec:temporal} focuses on the evolution of the user vocabularies over time.

The investigation of social ties between pairs of users focuses first on unveiling the characteristics of the implicit ties between users based on the similarity between their tagging activities (Section~\ref{sec:interest_sharing}). Additionally, this work explores the relationship between the strength of such implicit ties and those of more explicit social ties such as co-membership in discussion groups and semantic similarity of tag vocabularies (Section~\ref{sec:collaboration}). Studying the relationship among the implicit and explicit ties is relevant as we test whether the implicit ties based on usage similarity provide information about the potential creation of explicit social ties and ultimately for collaboration.

This study uses activity traces from three distinct tagging systems - {\em CiteULike}, {\em Connotea}, and {\em del.icio.us} (detailed in Section~\ref{sec:data}). We believe that this selection of systems samples the diversity of the tagging ecosystem, as they are three emblematic tagging systems for the type of content they target, with {\em CiteULike} and {\em Connotea} concentrating in bookmarking of academic citations, and {\em del.icio.us} focusing on general URLs. The in-depth analysis of these three systems reveals regularities and relevant variations in tagging behavior.

The main findings of this work are:

\begin{itemize}

  \item The characteristics of peer production of information are qualitatively similar across systems but differ quantitatively, as suggested by the observed rates of item re-tagging and tag reuse. In all three systems investigated, users produce new items at higher rate than they produce new tags. However, the observed rates in {\em CiteULike} and {\em Connotea} are different from {\em del.icio.us}.  As the three systems provide essentially similar annotation features, these findings suggest that the target audience and the type of annotated content play an important role in the users’ tagging behavior (Section~\ref{sec:reuse}).

  \item User tag vocabularies are constantly growing, but at different rates depending on the age of the user. However, despite the constant increase in size, the relative usage frequency of tags in a vocabulary converges to a stable ranking at early stages of a user's lifetime in the system. These observations have implications for applications that rely on tag-vocabulary similarity (e.g., recommender systems): these applications can use only a subsample of the entire user activity to estimate vocabulary similarity between users. Moreover, applications can aim to strike a balance between the accuracy of similarity estimates, the data volume used for estimation, and the freshness of the data. (Section~\ref{sec:temporal})

  \item The observed levels of activity similarity between pairs of users are the result of shared interested as opposed to generated by chance. The distributions of activity similarity strength deviate significantly from those produced by a Random Null Model (RNM) \cite{Reichardt2008}. This suggests that the implicit ties between users, as defined by their activity similarity levels, capture latent information about user relationships that may offer support for optimizing system mechanisms. (Section~\ref{sec:interest_sharing})

  \item The implicit social ties are related to explicit indicators of collaboration. We show that user pairs that share interests over items (i.e., annotate the same items) have higher similarity regarding the groups they participate together and higher semantic similarity of their tag vocabularies (even after eliminating the portions of tagging activity that is related to the items they tag in common). (Section~\ref{sec:collaboration}).

\end{itemize}

These characteristics have practical implications for the design of mechanisms that rely on implicit user interactions such as collaborative search~\cite{Evans2008,Yahia2008}, spam detection~\cite{Koutrika2008,Neubauer2009}, recommendation~\cite{Santos-Neto2007,Jaschke2007,Sigurbjornsson2008,Song2008} and the desig of incentives~\cite{Santos-Neto2010} as outlined in Section~\ref{sec:conclusions}.

\section{Related Work}
\label{sec:related_work}

This section contextualizes this work along four main topics: i) general characterization studies of peer production of information in tagging systems; ii) characterization of the evolution of tag vocabularies; iii) graph-based approaches to study activity similarity among users; and, iv) design of tag-based support mechanisms.

\subsection{General Characterization Studies}

Previous characterization studies focusing on tagging systems vary along three main aspects: {\em i}) the system analyzed from social bookmarking systems such as {\em del.icio.us}, {\em CiteULike}, and {\em Bibsonomy} to content sharing systems like {\em Flickr} and {\em YouTube}; and, {\em ii}) the focus of the characterization system-, tag-, item- or user-centric analysis; and, {\em iii}) the method of investigation - qualitative or quantitative research methods.
 
Nevertheless these works share the same intent: they address the high level set of questions that relate to characterizing the usage patterns observed and gaining insight into the underlying processes that generate them.  These works propose models that can be used to explain the observed characteristics of tagging activity such as the incentives behind tagging, the relative frequency of tags over time for a given item, the interval between tag assignments performed by users and the distributions of activity volume.

Hammond et al.~\cite{Hammond2005} is, perhaps, the first work to perform an initial stuy and to discuss the characteristics of social tagging, its potential, and the incentives behind tagging itself. The study comments on the features provided by different social tagging systems and discusses preliminary reasons that incentivize users to annotate and share content online. Following on the question of incentives, Ames et al.~\cite{Ames2007a} study tagging in online social media websites by interviewing 13 users on the fundamental question of {\em why do people tag}? Based on user answers, the authors suggest that tagging serves to support content organization or to communicate aspects about the content. These actions can be either socially- or personally-driven. More recent studies have followed the analysis of incentives at a larger scale~\cite{Strohmaier2012}. Our study supports and, more importantly, extends these result by performing a large-scale user behavior analysis (covering more than 700,000 users) in three tagging systems. Although, we do not focus on the question of incentives particularly, the quantitative analysis we present highlight and provide stronger evidence of existing incentives hypothesized by previous works.

One of the first works on the quantitative characterization of tagging systems is an item-centric characterization of {\em del.icio.us} that proposes the Eggenberger-Polya's urn model~\cite{Eggenberger1923} as an explanation to the observed relative frequencies of tags applied to an item~\cite{Golder2006}. Cattuto et al.~\cite{Cattuto2007} show in a tag-centric characterization that the observed tag co-occurrence patterns in {\em del.icio.us} is well modeled by the Yule-Simon's stochastic process~\cite{Simon1955}. Similarly, Capocci et al.~\cite{Capocci2009} show that the tag interarrival time distribution follows a power-law. Using a different approach, Chi and Mytkowicz~\cite{Chi2008} study the impact of user population growth in the efficiency of tags to retrieve items in {\em del.icio.us}. More recent works, focus on a characterization of social tagging systems that analyzes the impact of using tagging on external applications such as information retrieval and expert-generated content~\cite{Gu2011,Li2011,Lu2010,Seki2010}.

Another stream of characterization studies focuses on user-centric analysis. Nov et al.~\cite{Nov2008} present a user-centric qualitative study on the motivations behind content tagging in {\em Flickr}, where they suggest that users tag content due to a mixture of individual like personal content organization, and social motivation such as to help others in finding photos from a particular place. In a previous study, we characterize the user-centric properties of tagging activity from two social bookmarking systems designed for academic citation management: {\em CiteULike} and {\em Bibsonomy}. The observations suggest that user activity across the system follows the Hoerl model~\cite{Santos-Neto2007}.

Our work complements and extends these previous studies as it investigates a combination of user-, item- and tag-centric characteristics. Moreover, it explores different aspects of tagging activity, such as the levels of item re-tagging and tag reuse over time and the relationship between implicit and explicit user ties in tagging systems. By applying a quantitative approach on a broad population of users and multiple tagging systems, this study also offers new insights on user behavior that complement previous qualitative work by Ames and Naaman~\cite{Ames2007a}.

\subsection{Evolution of Users' Tag Vocabularies}

Tags represent to a certain extent the user perception or intended use of an item. It is natural, therefore, to assume that the set of tags (i.e, tag vocabulary) of a given user provides information about her topics of interest, which is useful to design other mechanisms that support efficient content usage such as recommender systems. Naturally, if tag vocabularies are stable over time, that is, if inclusion of new tags and shifts in the tag usage frequency observed in a vocabulary are rare, a mechanism can delay updates on the vocabulary snapshot used to base its predictions. Indeed, this study shows that this is the case (Section~\ref{sec:temporal}).

Previous studies on the characterization of the evolution of tag vocabulary can be divided in two categories: first, studies that aim to quantify and model the growth of tag vocabularies at both the system- and user-level~\cite{Cattuto2007,Cattuto2009}; and, second, studies that estimate shifts in the tag vocabularies over time such as evolution of the tag popularity distribution of item-level tag vocabularies~\cite{Halpin2007}, and the variation of tag usage frequency across predefined tag classes~\cite{Golder2006} (i.e., factual tags, subjective tags and personal tags)~\cite{Sen2006}.

In summary, these previous studies show that: {\em i}) the system-level and user-level tag vocabulary growth is sublinear; {\em ii}) item-level tag popularity distribution converges to a power-law; and, {\em iii}) the usage frequency of tag categories shifts over time.

This study extends previous works by evaluating different facets of the vocabulary evolution. First, this work goes beyond the estimation of vocabulary growth, focusing on the evolution of tag usage frequency. Second, it concentrates on individual, user-level tag vocabularies, as opposed to the item-level vocabularies as in the previous studies. Finally, it uses a different methodology to estimate the difference between tag vocabularies from different points in time. Finally, we note that we use a different approach that does not make assumptions about the categories of tags that appear in the user tag vocabularies, an approach used by previous works.

\subsection{Interest Sharing Analysis}

An alternative way to characterize tagging systems is a graph-centric approach. Two users are connected by a weighted edge with strength proportional to the similarity between the tagging activities of these two users. In this study, this similarity is referred to as an implicit social tie between users. Note that other types of connections between users are possible. In particular, we refer to explicit social ties as explicit indicators of user collaboration, such as co-membership in discussion groups.

This approach has been used by Iamnitchi et al.~\cite{Iamnitchi2011,Iamnitchi2004} to characterize scientific collaborations, the web, and peer-to-peer networks. The same model has been used by Li et al.~\cite{Li2011} to target the problem of finding users with similar interests in online social networking sites. The authors use a {\em del.icio.us} data set and define links between users based on the similarity of their tags. Their conclusions support the intuition that tags accurately represent the content by showing that tags assigned to a URL match to a great extent the keywords that summarize that URL. Additionally, they design and evaluate a system that clusters users based on similar interests and identifies topics of interests in a tagging community.

Another focus of graph-centric characterizations is to determine structural features in the graph formed by connecting users, items and tags based on similarity. Hotho et al.~\cite{Hotho2006} models a collaborative tagging system as a tripartite network (the network connects users, items and tags in a hypergraph) and design a ranking algorithm to enable search in social tagging systems. Using the same tripartite network model, Cattuto et al.~\cite{Cattuto2007} study {\em Bibsonomy} and show the existence of small-world patterns in such networks representing social tagging systems. Krause et al.~\cite{Krause2008a} also explore the topology of a tagging system, but the one formed by item similarity, to compare the folksonomy inferred from search logs and tagging systems. Their results suggest that search keywords can be considered as tags to URLs. More recently, Kashoob et al.~\cite{Kashoob2012} characterizes and model the temporal evolution of sub-communities in social tagging systems by looking into the similarity between users vocabularies.

Our study differs from these previous investigations in three aspects: first, the characterization of tagging activity similarity between users focuses on the system-wide concentration and intensity of pairwise similarities, as opposed to the topological characteristics. Second, our methodology provide a principled way to test whether the user similarity observed in social tagging systems is the product of interest sharing among users or chance. Finally, we investigate possible correlations between the observed levels of activity similarity between users (i.e., the implicit social ties) and the external indicators of explicit collaboration (i.e., the explicit social ties) as co-membership to discussion groups and semantic similarity of tag vocabularies (Sections~\ref{sec:interest_sharing} and~\ref{sec:collaboration}). We note that our methodology is inspired by a previous work by Reichardt and Bornholdt that studies the patterns of similarity of product preferences among buyers and sellers on eBay~\cite{Reichardt2008}.

\subsection{System Design}

System characterization work is primarily motivated by it potential impact on system design. Thus, several studies propose to exploit characteristics of tagging systems to improve mechanisms such as recommendation~\cite{Jaschke2007,Sigurbjornsson2008,Song2008}, spam detection~\cite{Koutrika2008,Krause2008a,Neubauer2009,Noll2009}, top-k querying techniques~\cite{Schenkel2008,Yahia2008}, and search and ranking~\cite{Hotho2006,Yanbe2007,Heymann2008}.

The present work adds to these studies by providing evidence that tagging activity can be useful to support such mechanisms. For example, the characteristics of vocabulary evolution, as presented in Section~\ref{sec:temporal}, can be used in the design of tagging systems in distributed platforms to adjust the frequency in which the user profiles are updated across nodes/users.
\section{Data Collection and Notation}
\label{sec:data}

This section describes the tagging systems analyzed as well as their respective activity traces collected and analyzed in this study, and introduces the basic notation used in the rest of this article.

We choose to analyze three tagging systems: {\em CiteULike}, {\em Connotea} and {\em del.icio.us}. The first two are designed to help users organize references to scientific publications, while the third is a social bookmarking tool for any type of URL.

The main reason to focus on these systems is their popularity. Additionally, studying systems that target different audiences enables a broader comparison between tagging systems that target a niche of web users such as the scientific community (i.e., {\em CiteULike} and {\em  Connotea}) and a system where any web user is a potential client (i.e., {\em  del.icio.us}). Furthermore, the characterizations of multiple classes of systems are complementary. Our intuition is that a study of more specialized tagging systems -- in this case, for managing academic publications -- may reveal social structures that are harder to identify in generic systems such as {\em  del.icio.us}.

{\em CiteULike}, {\em Connotea}, and {\em del.icio.us} target different types of content and users, though all three systems can be described in terms of the same abstract entities. In these systems each user maintains a library: a collection of bookmarked items that, for the systems we study, are either citation records linked to online articles or URLs to generic web pages. A user may assign tags to items in her library. Additionally, a user may also tag items in other user's public library. Tags may serve to group items, as a form of categorization, or to help find items in the future~\cite{Golder2006,Nov2008}. The tagging activity can be private (i.e., only the user who generated the tags and items can access these annotations) or public. The analysis presented in the next sections concentrates on the public portion of the activity. A user can see what (public) tags other users assigned to an item when she is tagging it, thus the user is able to reinforce the choice of tags as appropriate by repeating the tags previously assigned to that item.

In the case of {\em  CiteULike} and {\em  Connotea}, an item can be added to a user's library (an action often referred to as item posting) in three ways: i) browse popular scientific literature portals (e.g., ACM Portal, IEEE Explorer, arXiv.org) and use their features that automate item posting; ii) search for items already present in other users' libraries and add them to her own library; and, iii) post a new item manually. In {\em del.icio.us}, users can use automatic bookmarking features or manually bookmark URLs.

Table~\ref{tab:data_summary} presents a summary of the data sets used in this investigation. The {\em  CiteULike} and {\em  Connotea} data sets consist of all tag assignments since the creation of each system in late 2004 until January 2009. The {\em  CiteULike} dataset is available directly from its website. For {\em  Connotea}, we built a crawler that leverages {\em  Connotea}'s API to collect tagging activity since December 2004 (no earlier activity was retrieved). Finally, the {\em  del.icio.us} dataset is available at the website of a previous study by G\"orlitz et al.~\cite{Gorlitz2008}\footnote{http://www.tagora-project.eu/}.

Note that we do not have access to browsing or click traces. The traces analyzed in this work contain records that indicate when items are annotated with a given tag and who was the user, but the traces do not inform whether a tag is subsequently used by a user to navigate through the system, for example. The data sets are 'cleaned' to reduce sources of noise, such as the default tag 'no-tag' in {\em  CiteULike}, tags composed only of symbols and other tags like the automatically generated 'bibtex-import', which are clear outliers in the popularity distribution.

\begin{table}
\centering
\tbl{Summary of data sets used in this study}{
\label{tab:data_summary}
\begin{tabular}{lrrr}
                    \hline\noalign{\smallskip}
                   & {\em  CiteULike} & {\em  Connotea} & {\em  del.icio.us} \\
                    \noalign{\smallskip}\hline\noalign{\smallskip}
Activity Period    & 11/2004 -- 01/2009 & 12/2004 -- 01/2009 & 01/2003 -- 12/2006 \\\hline
\# Users           &             40,327 &             34,742 &            659,470 \\ 
\# Items           &          1,325,565 &            509,311 &         18,778,597 \\
\# Tags (distinct) &            274,982 &            209,759 &          2,370,234 \\
\# Tag Assignments &          4,835,488 &          1,671,194 &        140,126,555 \\
\noalign{\smallskip}\hline
\end{tabular}}
\end{table}

\paragraph{Notation.} The rest of this paper uses the following notation to formally refer to the entities that comprise tagging systems. A tagging system is composed of a set of users, items and tags, respectively denoted by $U$, $I$, $T$. The tagging activity in the system is a set of tuples $(u, i, w, t)$, where $u \in U$ is a user who tagged item $i \in I$ with tag $w \in T$ at time $t$. The activity of a user $u \in U$ can be characterized by $A_u$, $I_u$ and $T_u$, which are respectively the set of tag assignments performed by $u$, the set of items annotated, and the vocabulary or set of tags used by u. The user's activity from the beginning of the trace up to a particular point in time is denoted by $A_u(t_0,t)$, $I_u(t_0,t)$ and $T_u(t_0,t)$, where $t_0$ and $t$ are timestamps, $t_0$ represents the begin of the trace, and $t_0 \leq t$.

The next sections focus on the analysis of the traces described above, starting with the characteristics of peer production of information in these systems.

\section{Tag Reuse and Item Re-Tagging}
\label{sec:reuse}

Let a new item (or tag) be an item (or tag) that has never been used in an annotation in the tagging system. If users introduce new items and tags frequently, efficiently harnessing information based on collective action is difficult, if not impossible. This is so because in this case information about future user actions towards the annotation of an item or use of a tag is then hard to predict: prediction relies on the historical use of items and tags; new items or tags have no history in the system. Understanding the degree to which items are repeatedly tagged and tags reused can therefore help estimating the potential efficiency of techniques that rely on similarity of past user activity (e.g., recommender systems). To this end, this section addresses the following questions:
 
\begin{itemize}

  \item[Q1.1.] {\em What is the rate of repeated item annotation and tag reuse}? (Section~\ref{sec:levels})

  \item[Q1.2.] {\em Is the flow of new incoming users a major factor in the observed low rates of repeated item annotation}? (Section~\ref{sec:new_users})

  \item[Q1.3.] {\em Are the reuse patterns we observe the result of different usage characteristics of a group of high-volume {\em power} users, or are they pervasive through the entire user population}? (Section~\ref{sec:power})

\end{itemize}

The rest of this section first formalizes the metrics item re-tagging and tag reuse used to address these questions. Second, it characterizes the levels of item re-tagging and tag reuse as well as the level of activity generated by returning users. Finally, it discusses the implications of the usage characteristics discovered.

\subsection{Levels of Item Re-tagging and Tag Reuse}
\label{sec:levels}

An item is re-tagged (repeatedly tagged) if one or more users tag it again (with the same or different tags) after it was tagged for the first time. Similarly, a tag is reused if it appears in the trace more than once (for the same or different items) with different timestamps. We aim to determine which portion of the activity falls in these categories.

\begin{definition}
The level of item re-tagging during a time interval $[t_{f-1},t_f)$ is the ratio between the number of items tagged during that interval that have also been tagged in the past $[t_0,t_f)$ to the total number of items tagged during the interval $[t_{f-1},t_f)$, as expressed by Equation~\ref{eq:reuse}. Tag reuse, denoted by $tr(t_{f-1},t_f)$, is similarly defined.
\end{definition}

\begin{equation}
\label{eq:reuse}
ir(t_{f-1}, t_f) = \frac{|I(t_0,t_{f-1}) \cap I(t_{f-1},t_f)|}{|I(t_{f-1}, t_f)|}         
\end{equation}

We use this definition to determine the aggregate level of item re-tagging and tag reuse in {\em CiteULike}, {\em Connotea} and {\em del.icio.us}. Table~\ref{tab:reuse_summary} presents the median daily item re-tagging and tag reuse over the entire traces (i.e., the time interval $[t_{f-1} ,t_f)$ encompasses a day). The results show that {\em CiteULike} and {\em Connotea} have relatively low levels of item re-tagging while {\em del.icio.us} has a higher level of item re-tagging, yet all three systems present similarly high levels of tag reuse. We hypothesize that the observed difference in item re-tagging between {\em del.icio.us} and their counterparts in {\em CiteULike} and {\em Connotea} is due to the type of content users bookmark in each system (with URLs of any type in the former, and academic literature in the latter).

\begin{table}
\tbl{A summary of daily item re-tagging and tag reuse}{
\label{tab:reuse_summary}
\centering
\begin{tabular}{ccccc}
\hline\noalign{\smallskip}
          & \multicolumn{2}{c}{Re-Tagged Items} & \multicolumn{2}{c}{Reuse Tags} \\
            \noalign{\smallskip}\hline\noalign{\smallskip}
            & Median & Std. Dev. & Median & Std. Dev. \\\hline
{\em CiteULike}   & 0.15   & 0.07      & 0.84   & 0.12      \\ 
{\em Connotea}    & 0.07   & 0.06      & 0.77   & 0.21      \\
{\em del.icio.us} & 0.45   & 0.17      & 0.86   & 0.07      \\
\noalign{\smallskip}\hline
\end{tabular}}
\end{table}

\begin{figure}
  \centering
  \includegraphics[width=0.49\linewidth]{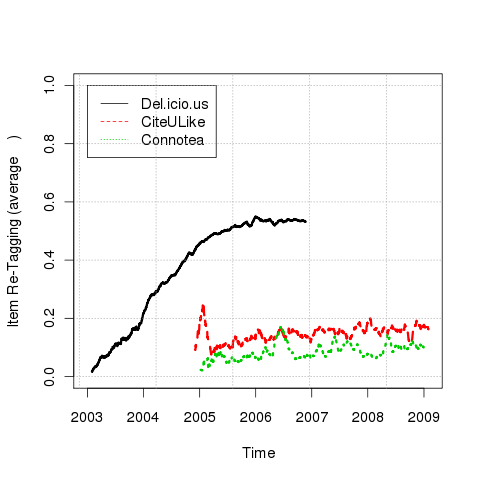}
  \includegraphics[width=0.49\linewidth]{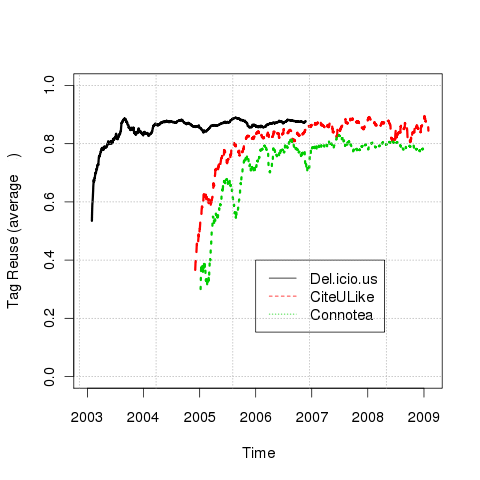}
  \caption{Daily item re-tagging (left) and tag reuse (right). The curves are smoothed by a moving average with window size $n=30$}
  \label{fig:reuse}
\end{figure}

To test whether these aggregate levels are a result of stable behavior over time, Figure~\ref{fig:reuse} presents the moving average (with a window size of 30 days) of daily item re-tagging and tag reuse. Overall, these results show that all three systems go through a bootstrapping period, after which they stabilize, with the levels of item re-tagging and tag reuse stabilizing much sooner for {\em CiteULike} and {\em Connotea} than that for {\em del.icio.us}. However, the tag reuse levels have a similar  evolution pattern in all three systems.

On the one hand, from the perspective of personal content management, the observed levels of item re-tagging and tag reuse, together with the much larger number of items than tags in these systems, suggest that users exploit tags as an instrument to categorize items according to, for example, topics of interest or intent of usage ('toread', 'towatch'). On the other hand, from the social (or collaborative) perspective, the relatively high level of tag reuse taken together with the low level of item reuse suggests that users may have common interest over some topics, but not necessarily over specific items. These quantitative results suggest that tags are used in the way previous exploratory qualitative study Ames and Naaman discusses~\cite{Ames2007a}.

A question that arises from the above observations is whether the levels of item re-tagging and tag reuse are generated by the same user or by different users. We observe that virtually none of the item re-tagging events are produced by the user who originally introduced the item to the system: generally, users do not add new tags to describe the items they collected and annotated once.

As illustrated by Figure~\ref{fig:self_reuse} (left), about $50\%$ of tag reuse is self-reuse (i.e., the reuse of a tag by a user who already used it first). This level of tag self-reuse indicates that users will often tag multiple items with the same tag, a behavior consistent with the use of tagging for item categorization and personal content management, as discussed above. Additionally, the fact that half of the tag reuse is not self-reuse  reinforces the notion that users do share tags, which indicates potentially similar interests. In Section 6, we further investigate this social aspect of tag reuse by defining and evaluating interest sharing among users, as implied by the similarity between users' activity (i.e., tags and items).
 
\subsection{New Incoming Users}
\label{sec:new_users}

\begin{figure}
  \centering
  \includegraphics[width=0.49\linewidth]{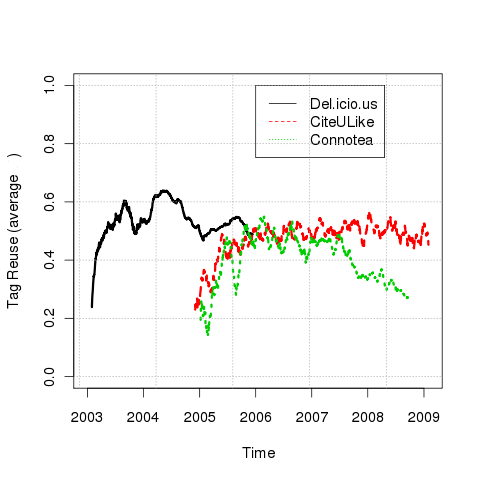}
  \includegraphics[width=0.49\linewidth]{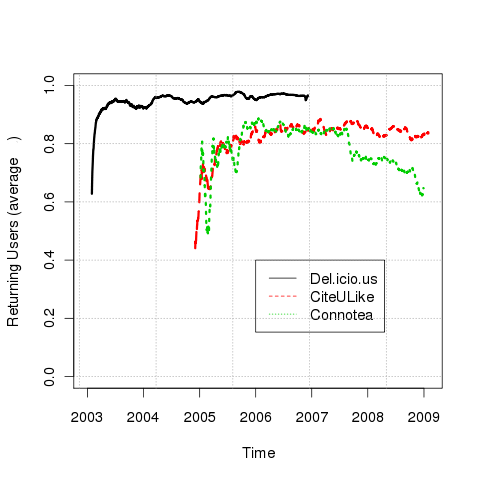}
  \caption{Self-tag reuse (left) and daily activity generated by returning users (right). The curves are smoothed by a moving average with window size $n=30$}
  \label{fig:self_reuse}
\end{figure}

To understand whether the observed low level of item re-tagging is due to a high rate of new users joining the community, we estimate the levels of activity generated by returning users (as opposed to new users that join the community). Figure~\ref{fig:self_reuse} (right) shows that, after a short bootstrap period, the level of tagging activity generated by returning users remains stable at about 80\% over the rest of the trace for both {\em CiteULike} and {\em Connotea}. In {\em del.icio.us}, the percentage of activity represented by returning users is even higher, with above 95\% of daily activity performed by returning users.
 
Thus, the low levels of item re-tagging are the outcome of expanding interests of returning users, instead of a constant stream of new users joining the community and introducing new items.

\subsection{The Influence of Power Users}
\label{sec:power}

Finally, we investigate the influence of highly active users in the observed item re tagging and tag reuse levels. To this end, we compare the observed item re-tagging and tag reuse with and without the activity produced by such power users. In this experiment, we define {\em power users} as the top-1\% most active users according to the number of annotations produced, and calculate item re-tagging and tag reuse as before.

The experiments test the hypothesis that the levels of item re-tagging and tag reuse are the same with and without the activity produced by these {\em power users}. To this end, we apply the Kolmogorov-Smirnov test (KS-test) on the two samples of activity (i.e., with and without the power users) with the null hypothesis that the item re-tagging and tag reuse observed in the two samples come from the same distribution (i.e, $H_0 =$ {\em the item re-tagging and tag reuse levels are equally distributed with and without the power users}). Using the KS-test is appropriate as it does not require that the samples are drawn from a normal distribution.

At a confidence level of $99\% (\alpha = 0.01, p = 1 - \alpha)$, we can reject the null hypothesis for all the systems, except the item re-tagging levels for {\em del.icio.us} (see the {\em p-values} in Table~\ref{tab:power_users}). This means that removing the activity produced by the power users leads to statistically different levels of item re-tagging and tag reuse as indicated by the D-statistic in Table~\ref{tab:power_users} (i.e., the maximum difference between the two distributions).

We hypothesize as the {\em del.icio.us} is a system that focuses on social bookmarking of URLs of any type (as opposed to be restricted to scientific articles in {\em CiteULike} and {\em Connotea}), removing the top 1\% most active users do not affect the observed levels of item re-tagging because some items will attract the attention of many other less active users. These users contribute, therefore, in large part for the observed levels of item re-tagging in {\em del.icio.us}.

\begin{table}
\tbl{The statistical test results reject the hypothesis that the item re-tagging and tag reuse observations with and without the power users are equal}{
\label{tab:power_users}
\centering
\begin{tabular}{ccc}
\hline\noalign{\smallskip}
            & \multicolumn{2}{c}{{\bf Re-Tagged Items}} \\
            \noalign{\smallskip}\hline\noalign{\smallskip}
            & D-Statistic & $p$-value $<$ \\\hline
{\em CiteULike}   & 0.03516     & $2.2 \times 10^{-16}$ \\ 
{\em Connotea}    & 0.1889      & $2.2 \times 10^{-16}$ \\
{\em del.icio.us} & 0.0475      & 0.0768                \\
            \noalign{\smallskip}\hline
            & \multicolumn{2}{c}{{\bf Reuse Tags}} \\
            \noalign{\smallskip}\hline\noalign{\smallskip}
            & D-Statistic & $p$-value $<$ \\\hline
{\em CiteULike}   & 0.2858 & $2.2 \times 10^{-16}$\\ 
{\em Connotea}    & 0.2132 & $2.2 \times 10^{-16}$\\
{\em del.icio.us} & 0.1371 & $3.23 \times 10^{-16}$\\
\noalign{\smallskip}\hline
\end{tabular}}
\end{table}

 

\subsection{Summary and Implications}

The observed user behavior impacts the efficiency of systems that rely on the inferred similarity among items, such as recommender systems. On the one hand, the relatively low level of item re-tagging suggests a highly sparse data set (i.e., attempting to connect users based on similar items will connect only few user pairs). A sparse data set poses challenges when designing recommender systems as they typically rely on the similarity of users based on their past activity to make recommendations.

On the other hand, the higher level of tag reuse confirms that analyzing tags has the potential to circumvent, or at least alleviate, the sparsity problem described above. The tags and users that relate to each item could not only serve to link items and build an item-to-item structure, but could also potentially provide semantic information about items. This information may help, for instance, to design better bibliography and citation management tools for the research community.

The results on analyzing the impact of power users in the observed levels of item re-tagging and tag reuse support two ideas: first, the notion that some users are instrumental on reducing the sparsity on tagging data sets (i.e., without power users, tags and items would be reused less, therefore potentially lesser items would be connected through tags and users). In fact, recommender systems benefit directly from the activity produced by such power users, as they can connect more items via repeated tag usage. Second, the role of power users differs from system to system, potentially due to effects of population size and diversity of interests. In the largest and most diverse system, we consider, reuse is a result of the activity of less active users rather than only power users.

Finally, despite the sparse data set problem, the fact that users tend to permanently add fresh content, as indicated by the low level of item re-tagging, implies that the approach proposed by Yanbe et al.~\cite{Yanbe2007} would be useful in a search portal for academic content. They suggest that content updated often in tagging systems can be used to improve the freshness and relevance of search results produced by a search engine. Portals for academic publications, such as Google Scholar, could exploit this fact to improve the freshness and relevance of their search results by using a combination of the PageRank ranking algorithm~\cite{Brin1998} and annotations from systems like {\em CiteULike}, {\em Connotea} and {\em del.icio.us}.

\section{Temporal Dynamics of Users' Tag Vocabularies} 
\label{sec:temporal}

The item re-tagging and tag reuse analysis presented in the previous section shows that users constantly produce new information in the system, by adding both new items to their libraries and tags to their vocabularies, though at different rates.

Although user tag vocabularies are constantly growing, it is unclear whether the growth rate is uniform over time. More importantly, vocabulary growth may or may not imply changes in the relative tag usage frequency by a given user. Changes in these frequency can indicate shifts in user interests over time.

To better understand these aspects of tagging activity, this section characterizes the temporal dynamics of user tag vocabularies. In particular, we study the rate of change of user vocabularies over time, as it quantifies the growth rate and changes in tag usage frequency for each user vocabulary. Overall, the objective is to answer the following question:

\begin{itemize}
  \item[Q2.] {\em How do users' vocabularies change over time}? 
\end{itemize}


To address this question, this section quantifies the evolution of user tag vocabularies by considering both their vocabulary growth and the tag usage frequency at different points in time. More specifically, the experiments first characterize the growth of user vocabularies, and, second, estimate the distance between tag vocabularies as expressed by the distance between snapshots of a user's vocabulary at various points in time and her final vocabulary. To take into account tag usage frequency the tags are ordered according to their frequency (i.e., the number of times the user annotated an item with the tag). 

We note that this investigation is different from, but complements, previous work~\cite{Kashoob2012,Cattuto2007,Halpin2007,Sen2006}: first, it performs a user-centric vocabulary analysis as opposed to a system-centric characterization; second, it studies both growth and change in vocabulary content in contrast to only one of the dimensions; and, finally, our characterization concentrates on the entire user population, as opposed to subcommunities of interests (as indicated by tags) or the evolution of such communities. Yet, the methodology we introduce in this study can be applied in addition to those proposed in previous works for a richer understanding of tag vocabulary evolution. The rest of this section describes the methodology applied to identify the vocabulary evolution, presents the results, and discusses its implications on system design.

\subsection{Methodology}

We introduce time in the definition of a user vocabulary by defining the tag vocabulary of a user $T_u(s,f)$ as the set of tags used within the tag assignment interval $[s,f]$. A particular case is $T_u(1,n)$ when $1$ and $n$ indicate the timestamps of the first and the last observed tagging assignment by user $u$, respectively. Thus $T_u(1,n) = T_u$ and represents the user's entire vocabulary.  

{\bf Vocabulary growth.} To analyze the vocabulary growth, we track the distribution of growth rates across the user population for the duration of the traces. The goal is to understand whether the growth rate changes according to the user age. Therefore, we measure the following ratio:

\begin{equation}
\frac{|T_u (1,k+1)| - |T_u (1,k)|}{|T_u (1,k+1)|}
\end{equation}

where $k \in [1, n]$ for all users in the system (i.e., $1$ and $n$ represent the timestamp of the first and last tag assignments of a particular users, respectively).

{\bf Vocabulary change.} To measure the rate of change in the content of the vocabularies, we consider vocabularies as sets of tags ordered in decreasing order of usage frequency (i.e., number of times the tag was used to annotate any item), and apply a distance metric as follows.

In this context, the final tag vocabulary, $T_u(1,n)$ is taken as a reference point to study the evolution of tag vocabularies in terms of the usage frequency of individual tags. The rationale behind the choice of this reference is that according to the tag reuse results in Section 4, user tag vocabularies are constantly growing. Therefore, it is unlikely that splitting the activity trace into disjoint windows could help identifying meaningful evolution patterns. Instead, we trace the evolution of a user's tag vocabulary by comparing the distance of incremental snapshots to her final vocabulary. This way, it is possible to understand the rate of convergence of user vocabularies over time. The experiment consists of calculating the distance from the tag vocabularies $T_u(1,k)$ ($k \in [2, n]$), to the reference tag vocabulary $T_u(1,n)$.

A traditional metric to calculate the distance between two lists of ordered elements is the Kendall's $\tau$ distance~\cite{Kendall1938}, which considers the number of pairwise swaps of adjacent elements necessary to make the lists similarly ordered. However, Kendall's $\tau$ distance assumes that both lists are composed of the same elements. Since we are interested in the evolution of tag vocabularies over time, this assumption is not valid in our case: tag vocabularies are likely to contain different tags at different times due to the constant inclusion of new tags.

Therefore, we apply the {\em generalized Kendall's $\tau$ distance}, as defined by Fagin et al.~\cite{Fagin2003}, which relaxes the restriction mentioned above and accounts for elements that are present in one permutation, but are missing in the other. Similar to the original Kendall's $\tau$ distance, the generalized version of the metric counts the number of pairwise swaps of items necessary to make the lists similarly ordered. Additionally, the generalized version counts the absence of items via a parameter $p$. This parameter can be set between 0 and 1, which allows various levels of certainty about the order of absent items. For example, in the case that two items are missing from one list, but present on the other, setting $p=0$ indicates that there are not enough information to decide whether the two items are in the same other or not. Conversely, setting $p=1$ indicates that there is full information available to consider the absence as an increase in the distance between the lists. In the experiments that follow we use $p=1$.

\subsection{Results and Implications}

Our analysis filters out users that had negligible activity considering only users with at least 10 annotations. This sample is responsible for approximately $93\%$, $61\%$, and $90\%$ of the total system activity in terms of tag assignments in {\em CiteULike}, {\em Connotea}, and {\em del.icio.us}, respectively.

\begin{figure*}
  \centering
  \includegraphics[width=0.31\linewidth]{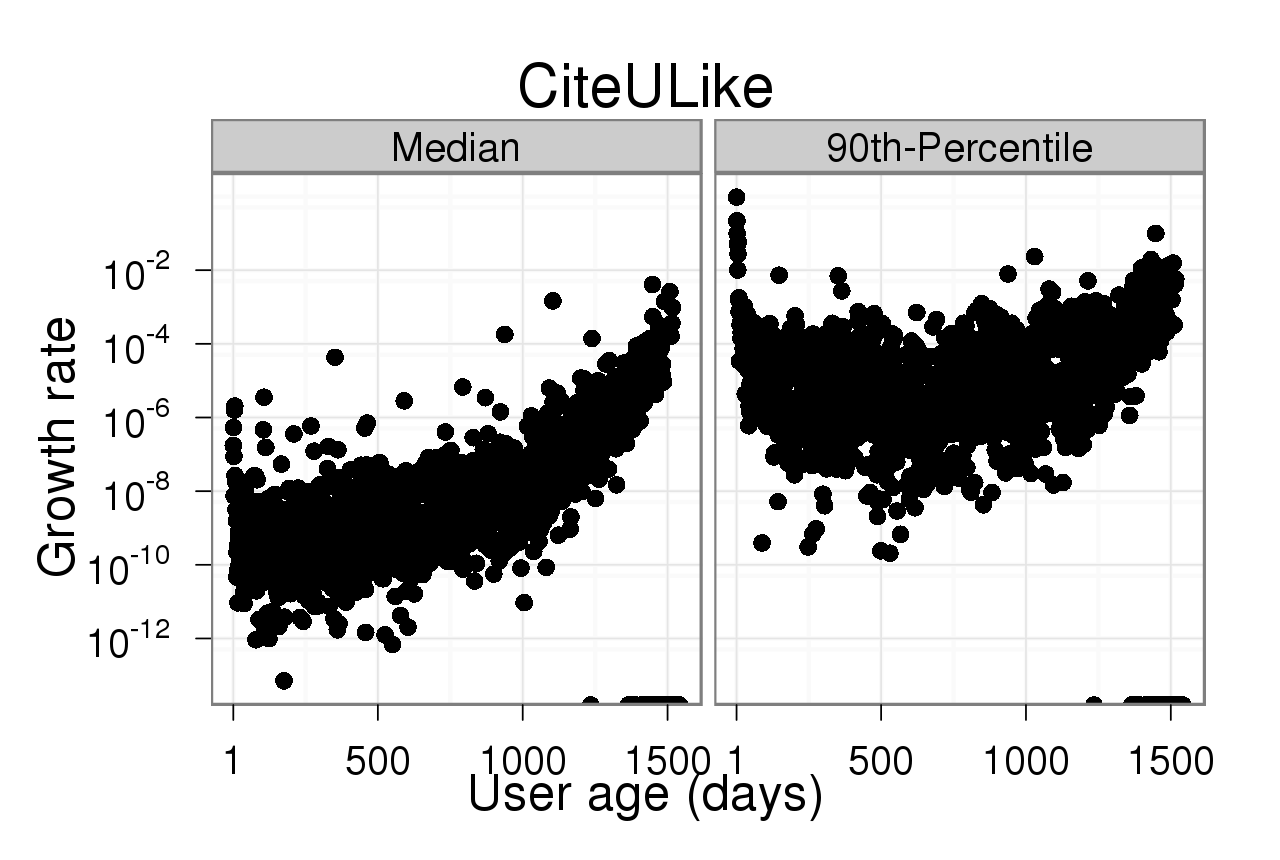}
  \includegraphics[width=0.31\linewidth]{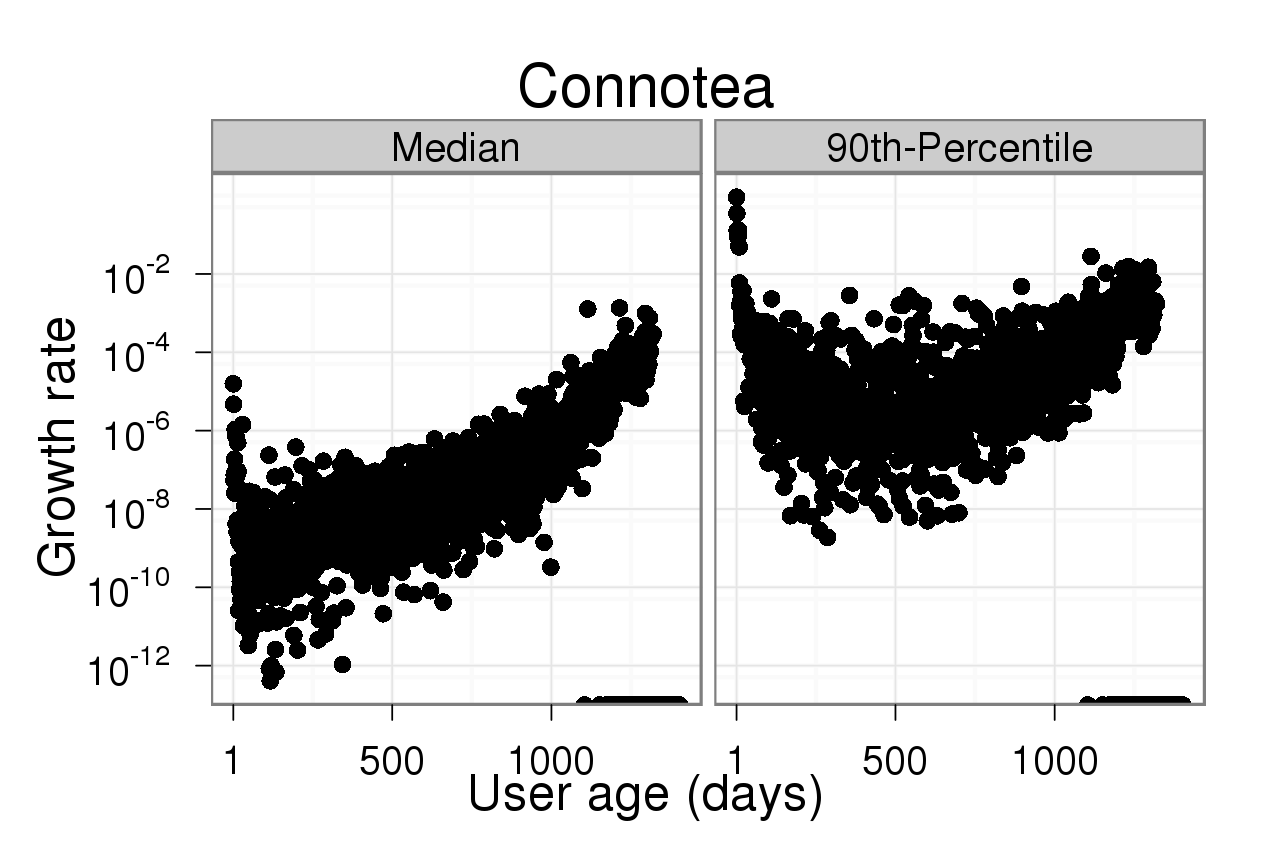}
  \includegraphics[width=0.31\linewidth]{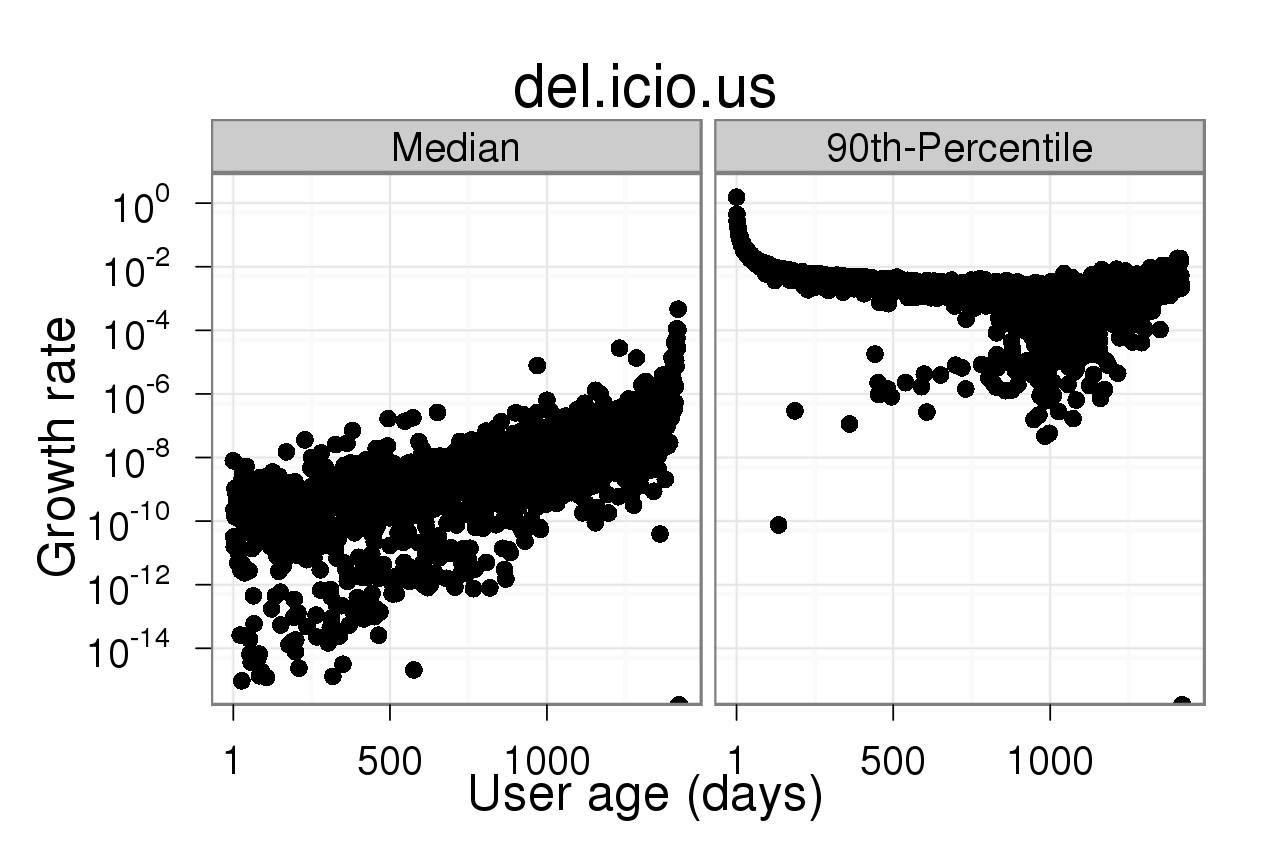}
  \caption{The vocabulary growth pattern in the systems studied: {\em CiteULike} (left), {\em Connotea} (center), and delicious (left)}
  \label{fig:vocabulary_growth}
\end{figure*}

{\paragraph Vocabulary growth rate.} Figure~\ref{fig:vocabulary_growth} illustrates vocabulary growth rate across the user population in the three systems studied. The x-axis indicates categories of users according to their age (i.e., number of days since their first recorded tag assignment), while the y-axis indicates the growth rate relative to each user vocabulary. For each of the systems studied we present two plots: labeled 'median' and '90th percentile'. A point in the median plot indicates that 50

The results show that, for the duration of the traces analyzed, the median growth rate (Figure~\ref{fig:vocabulary_growth} -- left) is relatively larger for older users. On the other hand, if we take the 90th percentile growth rate (Figure~\ref{fig:vocabulary_growth} -- right), except the very young users, we observe that the rate is relatively the same for all age groups with a slightly smaller rate for users in the middle of the age spectrum. An important observation is that except for the growth rate of young vocabularies, the $90^{th}$ percentile reaches a maximum rate of 0.1. This means that for $90\%$ of users, their vocabularies growth rate upper bound is $10\%$.   
   
\begin{figure*}
  \centering
  \includegraphics[width=0.31\linewidth]{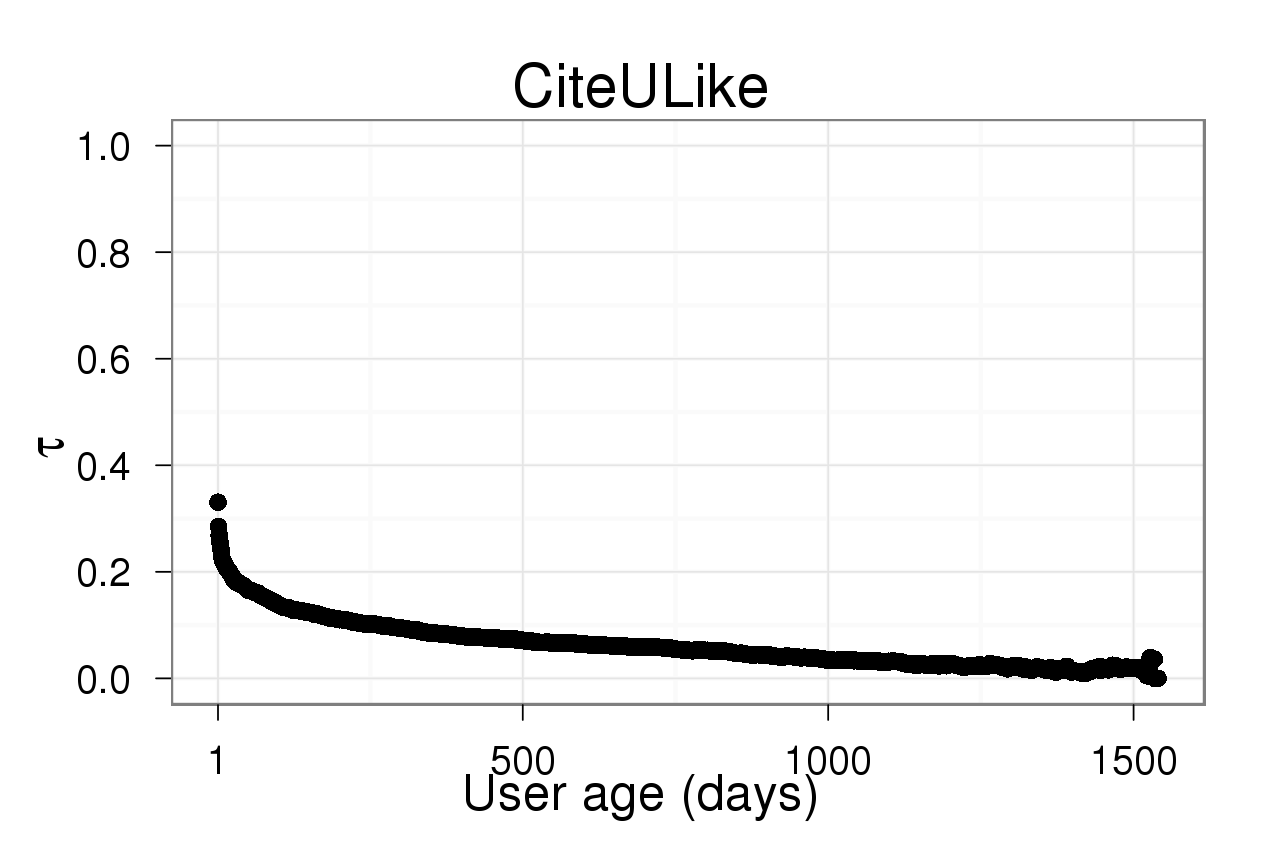}
  \includegraphics[width=0.31\linewidth]{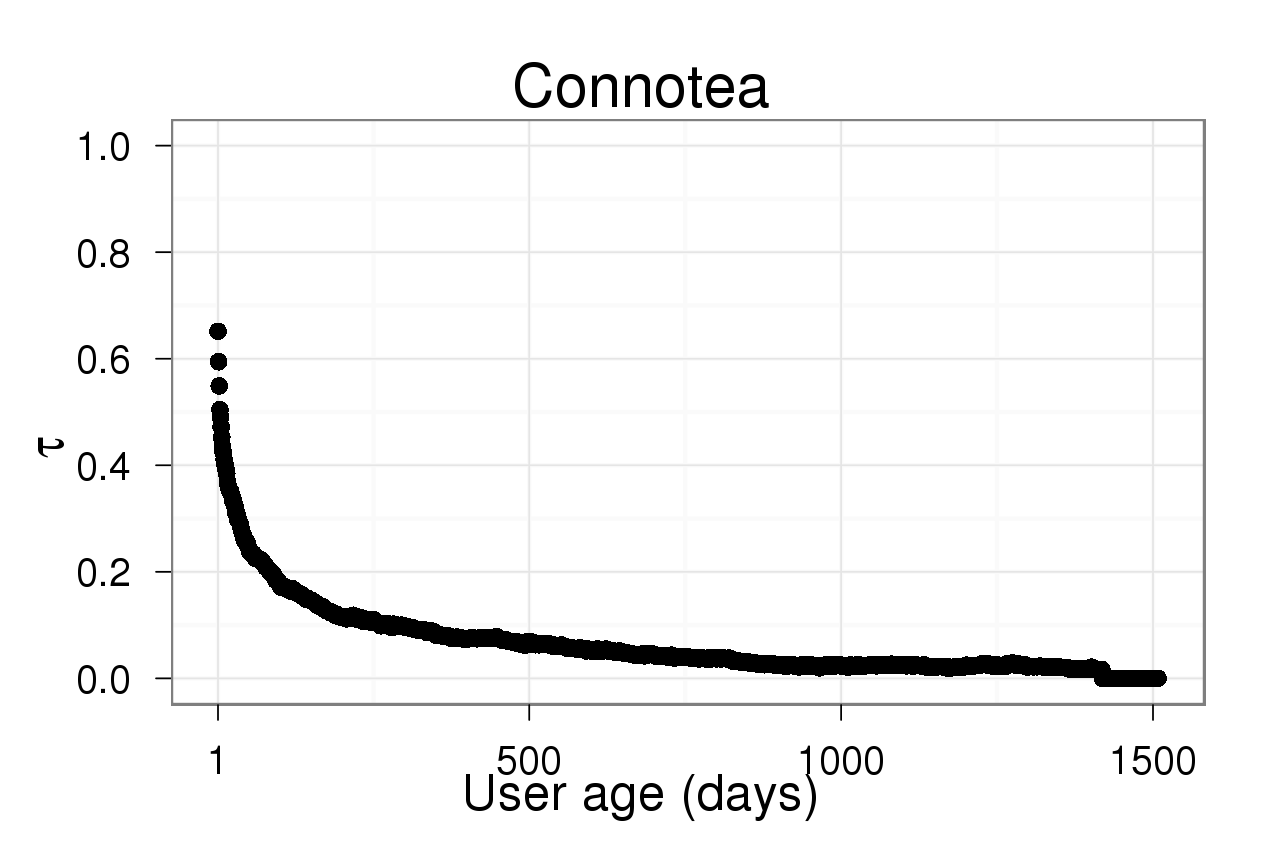}
  \includegraphics[width=0.31\linewidth]{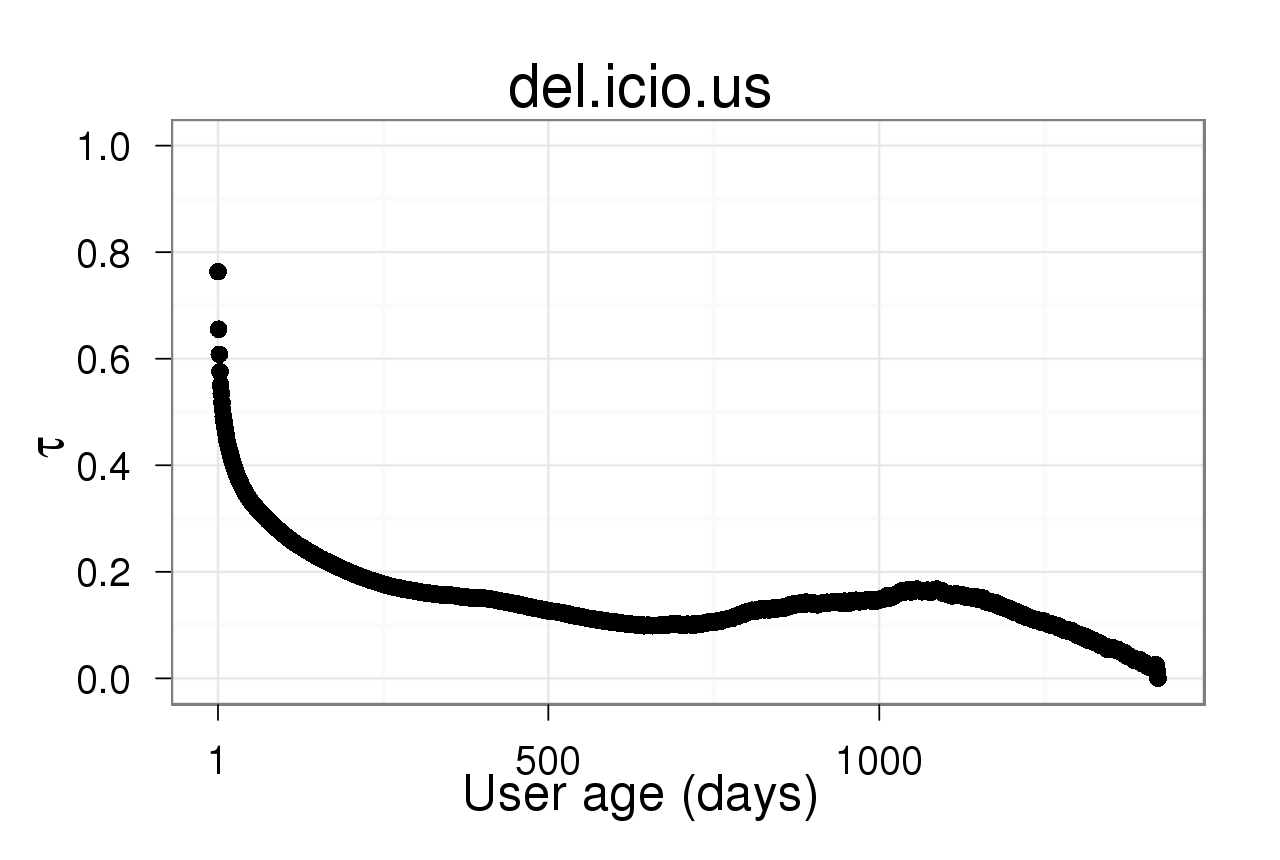}
  \caption{Rate of change in the tag usage frequency in the user vocabularies: {\em CiteULike} (left), {\em Connotea} (center), and {\em del.icio.us} (right)}
  \label{fig:vocabulary_change}
\end{figure*}

{\paragraph Vocabulary change.} Figure~\ref{fig:vocabulary_change} changes the focus from growth rate to the rate of change in users' vocabularies. The figure presents the rate of change in the contents of user vocabularies by taking into account the frequency of tags and calculating the distance between vocabulary snapshots. The results show that the distance from the vocabulary at earlier ages to its final state (i.e., Kendall-tau distance $t(T_u (1,k),T_u (1,n))$, where $k \in [2,n]$) decreases rapidly in the first 100 days for 50\% of users.


\section{Interest Sharing}
\label{sec:interest_sharing}

The analysis of item re-tagging and tag reuse in Section~\ref{sec:reuse} suggests that the observed level of re-tagging is the result of different users interested in the same item and annotating it. We dub this similarity in item related activity {\em item-based interest sharing}. Similarly, we dub the similarity in tag related activity {\em tag-based interest sharing}. This section defines and characterizes pairwise interest sharing between users as implied by their annotation activity in {\em CiteULike}, {\em Connotea} and {\em del.icio.us}.

Analyzing interest sharing is relevant for information retrieval mechanisms such as search engines tailored for tagging systems~\cite{Yahia2008,Zhou2008}, which can exploit pairwise user similarity to estimate the relevance of query results. This section focuses in particular on characterizing interest sharing distributions across the user-pairs in the system and addresses the following question:

\begin{itemize}
  \item[Q3.] {\em How is interest sharing distributed across the pairs of users in the system}? 
\end{itemize}

However, this section goes one step further and studies the system-wide characteristics of interest sharing and the implicit social structure that can be inferred from it. Moreover, the next section investigates the relationship between interest sharing (as inferred from activity similarity) and explicit indicators of collaboration such as co-membership in discussion groups and semantic similarity between tag vocabularies (Section~\ref{sec:collaboration}).

\subsection{Quantifying Activity Similarity}

We use the Asymmetric Jaccard Similarity Index~\cite{Jaccard1912} to quantity similarity between the item  (or tag-) sets of two users. We note that previous work (including ours) has used the Jaccard Index to quantify interest sharing: Stoyanovich et al.~\cite{Stoyanovich2008} used this index to model shared user interest in {\em del.icio.us} and to evaluate its efficiency in predicting future user behavior. Chi, Pirolli and Lam~\cite{Chi2007} applied the symmetric index to determine the diversity of users and its impact in a social search setting. Our analysis considerably extends that performed in previous work (as discussed in Section~\ref{sec:related_work}).

The formal definition of item-based interest-sharing metric is as follows (the tag-based version is defined similarly and denoted by $w_T$):

\begin{definition}
The level of item-based interest sharing between two users, $k$ and $j$, as perceived by $k$, is the ratio between the size of the intersection of the two item sets and the size of the item set of that user.
\end{definition}

\begin{equation}
\centering
\label{eq:interest_sharing}
w_I(k,j)=\frac{|I_k\cap I_j|}{|I_k|}
\end{equation}

Equation~\ref{eq:interest_sharing} captures how much the interests of a user $u_k$ match those of another user $u_j$, from the perspective of $u_k$. We opt for the asymmetric similarity index rather than the symmetric version (which uses the size of the union of the two sets as the denominator in Equation~\ref{eq:interest_sharing}) to account for the observation that the distribution of item set sizes in our data is heavily skewed. As a result, the situation where a user has a small item set contained in another user's much larger item set happens often. In such cases, the symmetric index would define that there is little similarity between interests, while the asymmetric index accurately reflects that, from the standpoint of the user with smaller item set, there is a large overlap of interests. From the perspective of the user with a large item set, however, only a small part of his interests intersect with those of the other user.

\subsection{How is Interest Sharing Distributed across the System?}
 
This section presents the distribution of pairwise interest sharing in {\em CiteULike}, {\em Connotea} and {\em del.icio.us}. We first find that approximately 99.9\% of user pairs in {\em CiteULike} and {\em del.icio.us} share no interest over items (i.e., $w_I(k,j) = 0$). In {\em Connotea}, the percentage is virtually the same: 99.8\%. For the tag-based interest sharing, the percentage of user pairs with no tag-based shared interest (i.e., $w_T(k,j) = 0$) is slightly lower: 83.8\%, 95.8\% and 99.7\% for {\em CiteULike}, {\em Connotea} and {\em del.icio.us}, respectively. Such sparsity in the pairwise user similarity supports the conjecture that users are drawn to tagging systems primarily by their personal content management needs, as opposed to the desire of collaborating with others.

The rest of this section focuses on the remaining user pairs, that is, those user pairs that have shared interest either over items or tags. To characterize these user pairs, we determine the cumulative probability distribution (CDF) of item- and tag-based interest sharing for these sets of user pairs in all three systems.

\begin{figure*}
  \centering
  \includegraphics[width=0.31\linewidth]{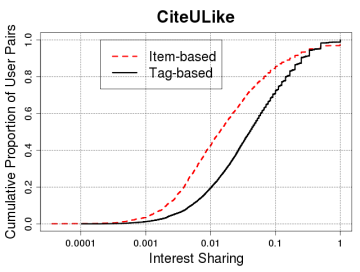}
  \includegraphics[width=0.31\linewidth]{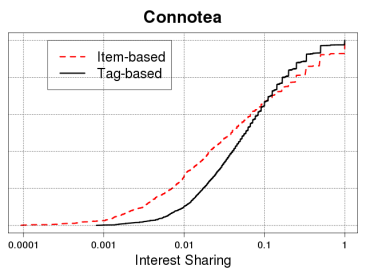}
  \includegraphics[width=0.31\linewidth]{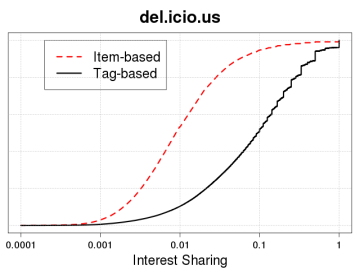}
  \caption{Distributions for item- and tag-based interest sharing (for pairs of users with non-zero sharing) in {\em CiteULike}, {\em Connotea} and {\em del.icio.us}}
  \label{fig:distribution}
\end{figure*}

Figure~\ref{fig:distribution} shows that, in all three systems, the typical intensity of tag-based interest sharing is higher than its item-based counterpart. This is not surprising: after all, all three systems include two to three times more items than tags. However, there is qualitative difference across systems with respect the concentration of item-based and tag-based interest sharing levels, with {\em del.icio.us} showing a much wider gap between the distributions.

The difference between the levels of item- and tag-based interest sharing suggests the existence of latent organization among users as reflected by their fields of interest. We hypothesize that this observation is due to a large number of user pairs that have similar tag vocabularies regarding high-level topics (e.g., computer networks), but have diverging interests in specific sub-topics (e.g., internet routing versus firewall traversal techniques), which could explain the relatively lower item-based interest sharing compared to the observed tag-based interest sharing.

Finally, to provide a better perspective in the tag-based interest sharing levels, we compare the observed values to that of controlled studies on the vocabulary of users describing computer commands [16]. The tag-based interest sharing level, as observed in Figure 6, is approximately 0.2 (or less) for 80\% of the user pairs that have some interest sharing, while Furnas et al. [16] show that in an experiment where participants are instructed to provide a word to name a command based on its description such that it is an intuitive name and more likely to be understood by other people, the ratio of agreement between two participants is in the interval [0.1, 0.2] (i.e., number of times two participants use the same word divided by the total number of participant pairs).

These observations suggest that observed tag-based interest sharing is due to conscious choice of terms from vocabularies that are shared among users, rather than by chance. We look more closely into this aspect in the next section by constructing a baseline to compare the observed interest sharing levels to that of a random null model.

\subsection{Comparing to a Baseline}

The goal of this section is to better understand the interest sharing levels we observe. In particular, we focus on the following high-level question:

\begin{itemize}
  \item[Q4.] {\em Do the interest sharing distributions we observe differ significantly from those produced by random tagging behavior?}
\end{itemize}

For this investigation, we compare the observed interest sharing distribution to that obtained in a \textit{system with users that have an identical volume of activity and the same user-level popularity distributions for items or tags, but do not act according to their personal interests}. Instead, in the random null model (RNM)~\cite{Reichardt2008}, the chance that a user is interested in an item or tag is simply that item or tag's popularity in the user's vocabulary.

The reason to perform this experiment is the following: we aim to validate our intuition that the interest sharing metric distils useful user behavior information. If the interest-sharing levels we observe in the three real systems at hand are more concentrated than those generated by the RNM, then interest sharing metric captures relevant information about similarity of user preferences, rather than simply coincidence in the tagging activity.

To reiterate, the random null model (RNM) is produced by emulating a tagging system activity that preserves the main macro-characteristics of the real systems we explore (such as the number of items, tags, and users, as well as item and tag popularity, and user activity distributions), but where users make random tag assignments. As such, random assignments are used here as the opposite of interest-driven assignments.

To test our hypothesis, we compare the two sets of data (real and RNM-generated) in terms of the numbers of user pairs with non-zero interest sharing and the interest-sharing intensity distribution. Because of its probabilistic nature, we use the RNM to generate five synthetic traces corresponding to each of the real systems we analyze. For the rest of this section, the RNM results represent averages over the five RNM traces for each system. We confirmed that the five synthetic traces represent a large enough sample to guarantee a narrow 95\% confidence interval for the average interest sharing observed from the RNM simulations.

Our data analysis shows that the observed interest sharing deviates significantly from that generated by random behavior in two important respects.

First, interest sharing (and, consequently, the similarity between users) is more concentrated in the real systems than in the corresponding simulated RNM. More specifically, the number of user pairs that share some item-based interest (i.e., $w_I(k,j) > 0$) is approximately three times smaller in the real systems than in the RNM-generated ones. Tag-based interest sharing follows a similar trend.

\begin{figure*}
  \centering
  \includegraphics[width=0.49\linewidth]{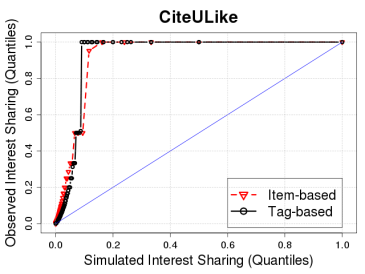}
  \includegraphics[width=0.49\linewidth]{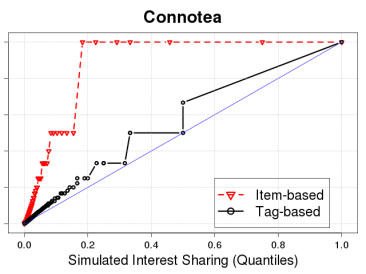}
  \caption{Q-Q plots that compare the interest sharing distributions for the observed vs. simulated (i.e., the RNM model) for {\em CiteULike} (left) and {\em Connotea} (right)}
  \label{fig:rnm}
\end{figure*}

Second, interest sharing distribution deviates significantly from that produced by a RNM. We compare the cumulative distribution function (CDF) for the interest sharing intensity for the user-pairs that have some shared interest (i.e., $w(k,j) > 0$). Figure~\ref{fig:rnm} presents the Q-Q plots that directly compare the quantiles of the distributions of interest-sharing levels derived from the actual trace and those derived from the simulated RNM. A deviation from the diagonal indicates a difference between these distributions: The higher the points are above the diagonal, the larger the difference between the observed interest-sharing levels and those generated by the RNM. 
  
We note that the only interest-sharing distribution that is close to the one produced by the RNM is for {\em Connotea}'s tag-based interest sharing (Figure~\ref{fig:rnm}). However, there is still a significant deviation from randomness: the real activity trace leads to three times fewer user-pairs that share interest than the corresponding RNM.

\subsection{Summary and Implications}

This section provides a metric to estimate pairwise interest sharing between users, offers a characterization of interest-sharing levels in {\em CiteULike} and {\em Connotea}; and investigates whether the observed interest sharing in these systems deviates from that produced by chance, given the amount of activity users had. Such reference is given by a random null model (RNM) that preserves the macro characteristics of the systems we investigate, but uses random tag assignments.

The comparison highlights two main characteristics of the interest sharing: first, interest sharing is significantly more concentrated in the real traces than in the RNM-generated activity: in quantitative terms, three times fewer user pairs share interests in the real traces. Second, most of the time, for the user pairs that have non-zero interest sharing the observed interest-sharing intensity is significantly higher in each real system than in its RNM equivalent.

We conjecture that a possible explanation for these observations is as follows. Let us consider that the set of tags that can be assigned to an item is largely limited by the set of topics that item is related to. In this case, intuitively, the probability of choosing a tag is conditional to the set of topics the item is related to. At one extreme, the maximum diversity of topics occurs when there is a one-to-one mapping between topics and tags, that is, when each tag introduces a different topic. The RMN simulates the other extreme, a single topic that encompasses all tags in the system.

However, in real systems, the interests for each individual user are limited to a finite set of topics, which is likely to determine their tag vocabulary. This leads to a concentration of interest sharing, as implied by the tag similarity, on few user pairs, yet at higher intensity than that produced by the RNM.

Finally, and most importantly, the divergence between the observed and the RNM-generated interest sharing distributions shows that activity similarity, our metric to quantify interest sharing intensity, embeds information about user self-organization according to their preferences. This information, in turn, could be exploited by mechanisms that rely on implicit relationships between users. The next section seeks evidence about the existence of such information by analyzing the relationship implicit user ties, as inferred from the similarity between users' activity, and their explicit social ties, as represented by co membership in discussion groups or semantic similarity between tag vocabularies.
\section{Shared Interest and Indicators of Collaboration}
\label{sec:collaboration}
 
The previous section characterizes interest sharing across all user pairs in each system and suggests that it encodes information about user behavior, as its distribution deviates significantly from that produced by a random null model.

This section complements this characterization and evaluates whether the implicit user relationships that can be derived from high levels of interest sharing correlate with explicit online social behavior. More specifically, this section addresses the following question:

\begin{itemize}
  \item[Q5.] {\em Are there correlations between interest sharing and explicit indicators of social behavior}? 
\end{itemize}

Before starting the analysis, it is important to mention that the number of externally observable elements of user behavior to which we have access is limited by the design of the tagging systems themselves (e.g., the tagging systems collect limited information on user attributes) and by our limited access to data (e.g., we do not have access to browsing traces or search logs).

One {\em CiteULike} feature, however, is useful for this analysis: {\em CiteULike} allows users to explicitly declare membership to groups and to share items among a selected subset of co-members -- an explicit indicator of user collaboration in the system. Thus, this feature enables an investigation about the relationship between interest sharing and group co-membership (which we assume to indicate collaboration). We note that a similar experiment could be performed using the explicit friendship links in {\em del.icio.us}, for example. However, this data is not available to our study.

Along the same lines, we use a second external signal: semantic similarity between tag vocabularies. More specifically, we test the hypothesis that item-based interest sharing relates to semantic similarity between user vocabularies. The underlying assumption here is that users who (have the potential to) collaborate employ semantically similar vocabularies.

This section presents the methodology and the results of these two experiments that mine the relationship between interest sharing and indicators of collaboration. In brief, our conclusions are:

\begin{itemize}
  \item User pairs with positive item-based interest sharing have a much higher similarity in terms of group co-membership and semantic tag vocabulary, than users who have no interest sharing.

  \item On the other side, we find no correlation between the intensity of the interest sharing and the collaboration levels as implied by group co-membership or vocabulary similarity.
\end{itemize}

\subsection{Group Membership}

In {\em CiteULike}, approximately 11\% of users declare membership to one or more groups. While the percentage may seem small, they are the most active users: these users generate 65\% of tag assignments, and introduce 51\% of items and 50\% of tags. For this section we limit our analysis to the user pairs for which both users are members of at least one group. Also, the analysis focuses on groups that have two or more users (about 50\% of all groups) as groups with only one user are obviously not representative of potential collaboration.

The goal is to explore the possible relationship between item-based interest sharing and co-membership in one or more groups. Let $H_u$ be the set of groups in which the user $u$ participates. We determine the group-based similarity $w_H(u,v)$ between two users $u$ and $v$ using the asymmetric Jaccard index, similar to the item-based definition in Eq.~\ref{eq:interest_sharing}, but considering the sets of groups users participate in. Based on this similarity definition, we study whether the intensity of item-based interest sharing between two users with non-zero interest sharing (i.e., $w_I(u,v) > 0$) correlates with group membership similarity.

We find no correlation between $w_I(u,v)$ -- the item-based similarity -- and $w_H(u,v)$ -- the group-based similarity. More precisely, Pearson's correlation coefficient is approximately $0.12$, and Kendall's $\tau$ is about $0.05$. This is surprising as one would expect that being part of the same discussion groups is a good predictor to the intensity in which users share interest over items. Therefore, we look into these correlations in more detail.

To put these correlation results in perspective, we look at group similarity for two distinct groups of user pairs: those with no item-based interest sharing ($w_I(u,v) = 0$) and those with some interest sharing ($w_I(u,v) > 0$). We observe that, although the group information is relatively sparse, pairs of users with positive interest sharing are more likely to be members of the same group than the user pairs where $w_I(u,v) = 0$. In particular, 4\% of the user pairs with $w_I(u,v) > 0$ have $w_H(u,v) > 0.2$, while twenty times fewer user pairs with $w_I(u,v) = 0$ have $w_H(u,v) > 0.2$.

These observations suggest that activity similarity is a necessary, but not sufficient condition for higher-level collaboration, such as participation in the same discussion groups. Although users share interest over items, and may implicitly benefit from each other tagging activity (e.g., using one another's tags to navigate the system), this may not directly lead to users actively engaging in explicit collaborative behavior. Conversely, the lack of interest sharing strongly suggests a lack of collaborative behavior.

\subsection{Semantic Similarity of Tag Vocabularies}

This section complements the previous analysis on the relationship between item-based interest sharing and collaboration indicators via group co-membership. It investigates the potential relation between item-based interest sharing of a pair of users and the semantic similarity between their tag vocabularies, that is, the set of tags each has applied to items in its library. Since, through this experiment we aim to understand the potential for user collaboration through similar vocabularies, when comparing vocabularies for a user pair, we exclude the tags applied to the items the two users have tagged in common -- a these tags have a likely high similarity.

The rest of this section is organized as follows: it presents the metric used to estimate the semantic similarity of two tag vocabularies; discusses methodological issues; and, finally, presents the evaluation results.

Estimating semantic similarity: We use the lexical database {\em WordNet} to estimate the semantic similarity between individual tags. {\em WordNet} consists of a set of hierarchical trees representing semantic relations between word senses such as synonymy (the same or similar meaning) and hypernymy/hyponymy (one term is a more general sense of the other). Different methods have been implemented to quantify semantic similarity using {\em WordNet}. In particular, WordNet::Similarity -- a Perl module -- provides a set of semantic similarity measures~\cite{Pedersen2004}.

For our experiments, we use the Leacock-Chodorow similarity metric~\cite{Budanitsky2006}, as previous experiments, based on human judgments, suggest that it best captures the human perception of semantic similarity. The metric is derived from the negative log of the path length between two word senses in the {\em WordNet} "is-a" hierarchy, and is only usable between word pairs where each has at least one noun sense.

Additionally, we explore a method to extend coverage to a larger subset of users' tag vocabularies, with an approach that builds on the {\sc yago} ontology, developed and described by Suchanek et al.~\cite{Suchanek2007,Suchanek2008}. {\sc yago} ("Yet Another Great Ontology") is built from the entries in Wikipedia~\footnote{http://wikipedia.org}, a collaborative online encyclopedia. The standardized formatting of Wikipedia makes it possible for information to be automatically extracted from the work of thousands of individual contributors and used as the raw material of a generalized ontology. The primary content of the {\sc yago} ontology is a set of fact tables consisting of bilateral relations between entities, such as "bornIn", a table of relations between persons and their birthplaces. Five of the relations are of particular interest to us because they contain links between entities mentioned in Wikipedia and terms found in {\em WordNet}. In this way, we are able to identify some tags as probable personal, collective, or place names, and use the {\em WordNet} links from {\sc yago} to map these on to a set of corresponding {\em WordNet} terms.

We used a merged tag vocabulary from {\em CiteULike} and {\em Connotea} datasets. A little over 13\% of the tags in the merged repertoire had direct matches in {\em WordNet}. By adding the tags matched through comparison with {\sc yago}'s {\em WordNet} links, this was increased to 28.6\% of unique tags applied by users of both systems. Note, however, that these tags cover up to 75\% of the tagging activity in the two systems, as shown in Table~\ref{tab:activity}.

\begin{table}
\tbl{The share of tagging activity captured by the tag vocabularies in {\em CiteULike} and {\em Connotea} that is found in {\em WordNet} and WorldNet combined with {\sc yago} lexical databases. As we use an anonymous version of the {\em del.icio.us} dataset, with all the users, items, and tags identified by numbers, this precluded us to perform the same analysis using {\em WordNet} and {\sc yago} for {\em del.icio.us}.}{
\label{tab:activity}
\centering
\begin{tabular}{ccc}
\hline\noalign{\smallskip}
          & {\em WordNet} only & {\em WordNet} + {\sc yago} \\
          \noalign{\smallskip}\hline\noalign{\smallskip}
{\em CiteULike} & 62.1\%  & 79.5\%  \\ 
{\em Connotea}  & 51.3\%  & 65.3\%  \\
Combined  & 57.4\%  & 73.4\%  \\
\noalign{\smallskip}\hline
\end{tabular}}
\end{table}

In order to match tags gathered from the two systems with corresponding entities in {\sc yago}, we replaced all non-ASCII characters, such as accented letters, with their nearest ASCII equivalents; removed all characters other than letters and numerals, and reduced all the {\sc yago} entities to lower case (tags from both systems being already reduced to lower case). We allowed partial matches, but required that the end of a tag correspond to a word boundary in the {\sc yago} entity or vice-versa. Thus, we were able to develop a mapping between about 58,600 tags from the merged vocabulary and 57,900 distinct {\em WordNet} senses, with most tags matching multiple {\em WordNet} senses. Given that the addition of {\em WordNet} terms identified by mapping through {\sc yago} effectively increases the total depth of the tree being considered, the Leacock-Chodorow algorithm required that we adjust all tag pair similarity scores accordingly in order to fairly compare the {\em WordNet}-only and {\em WordNet}+{\sc yago} scores. The maximum possible similarity with {\em WordNet} alone is $\log(1/40)$ or $3.689$; whereas with {\em WordNet} + {\sc yago} it is $\log(1/42)$ or $3.738$.

We define the similarity $sim(t_1,t_2)$ between two tags $(t_1, t_2)$ as the maximum Leacock-Chodorow similarity between every available noun sense of $t_1$ and $t_2$. Thus, we define the semantic similarity between the tag vocabularies $T_u$ and $T_v$ of two users, $u$ and $v$, as perceived by $u$, is denoted by $s(u,v)$, and determined by the ratio between the sum over the pairwise tag similarities and the size of $u$'s vocabulary, as expressed by Eq.~\ref{eq:sim} below.

\begin{equation}
\label{eq:sim}
\centering
sim(u, v) = \frac{\sum_{t_1 \in T_u, t_2 \in T_v} sim(t_1,t_2)}{|T_u|}
\end{equation}

We then calculate the corresponding value of $s(v,u)$ by reversing the $u$ and $v$ terms in Eq.~\ref{eq:sim} and record the smaller of the two -- i.e. $min(s(u,v), s(v,u))$ -- as the undirected tag vocabulary similarity between the two users $u$ and $v$. We note that this metric is based on the Modified Hausdorff Distance (MHD)~\cite{Dubuisson1994}.

\paragraph{Methodological issues.} There are three practical issues regarding our experimental design that deserve a note. First, to avoid bias, if two users assigned the same tags to the same item, we omit these tags from their vocabularies, before determining the aggregate similarity. By eliminating from vocabularies the tags that have been used on exactly the same items, we eliminate the tags on which the two users have most likely already converged. We look only at the remaining parts of the vocabularies where convergence is not apparent. Second, the Leacock-Chodorow similarity metric only considers words that have noun senses in {\em WordNet}, because it is calculated from paths through the "is-a" hierarchy, only defined for nouns. Tags in both systems considered may include words or phrases from any language, abbreviations, or even arbitrary strings invented by the user, while {\em WordNet} consists mainly of common English words. A third methodological issue was that matching tags to {\sc yago} entries, in some cases, returned an unmanageably large set of distinct {\em WordNet} senses. We accordingly eliminated those tags that were above the 99th percentile in distinct {\em WordNet} senses matched, which were those returning more than 167 distinct senses.
  
\begin{figure*}
  \centering
  \includegraphics[width=0.49\linewidth]{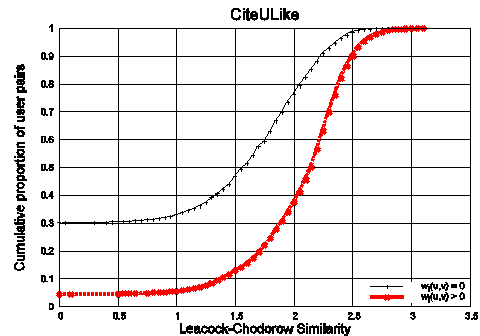}
  \includegraphics[width=0.49\linewidth]{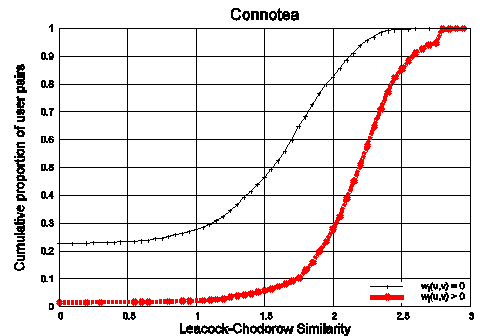}
  \caption{CDFs of tag vocabulary similarity for user pairs with positive (bottom curve) and zero (top curve) activity similarity. {\em CiteULike} (left); {\em Connotea} (right)}
  \label{fig:cdf_wordnet_yago}
\end{figure*}

\paragraph{Results.} We use sampling to test, in both {\em CiteULike} and {\em Connotea}, whether there is a significant difference in tag vocabulary similarity between two sets of user pairs: one where all users have no item-based interest sharing and one with positive item-based interest sharing (we sample each group with $n=4000$ pairs). This analysis shows that the vocabularies of user pairs with interest sharing are significantly more similar than those of user pairs with no interest sharing (Figure~\ref{fig:cdf_wordnet_yago}). The median vocabulary similarity for user pairs with positive interest sharing $\mu_c = 2.112$ ($\pm 0.02$, $99\%$ c.i.) is about 1.6 times that of user pairs with no interest sharing $\mu_u = 1.308$ ($\pm 0.04$, $99\%$ c.i.). This salient difference in the vocabulary similarity suggests that the item-based interest sharing embeds information about the "language" shared by the users to describe the items they are interested in.

\subsection{Summary and Implications}

This section takes a first step towards understanding the relationship between the implicit user ties, as inferred from pairwise interest sharing, and their explicit social ties. First, we look at correlations between the item-based interest sharing and the group-based similarity. The observations indicate that although the intensity of item-based activity interest sharing does not correlate with explicit collaborative behavior, as implied by group co-membership, user pairs with some interest sharing are more than one order of magnitude more likely to participate in similar groups.

Second, we evaluate the relationship between item-based interest similarity and the semantic similarity of tag vocabularies. We discover that, although the two do not yield a Pearson's correlation, item-based interest similarity does embed information about the expected semantic similarity between user vocabularies.

These results have implications on the design of mechanisms that aim to predict collaborative behavior, as these mechanisms could exploit item-based similarity to set expectations about group-based and vocabulary-based similarity. Moreover, assuming that the tagging activity characteristics of spammers differ from legitimate users, one could use deviations from observed relationship between item-based similarity and the two indicators of collaborative behavior presented here to detect malicious user behavior.

\section{Conclusions}
\label{sec:conclusions}

Tagging systems have been widely adopted by today's World Wide Web. These systems provide users with the ability to annotate and share content. The peer produced annotations (or tags) and shared items create a valuable pool of metadata for mechanisms that aim to harness or predict user preferences such as recommendation systems and social search. However, to efficiently use this information, it is first necessary to understand the usage characteristics of tagging systems.

To this end, this work studies two major aspects of usage characteristics in tagging systems: {\em i}) the dynamics of peer production of information; and, {\em ii}) the relationship between implicit and explicit social ties between users.

To address the first aspect, this work analyzes the user behavior characteristics at the individual and aggregate level in three tagging systems that focus on distinct applications: {\em CiteULike} and {\em Connotea} -- personal management of academic citation records; and, {\em del.icio.us} -- a popular social bookmarking system.

In particular, the characterization of peer production of information focuses on three user activity indicators: {\em i}) item re-tagging, a measure for the degree to which users re-tag the items already existing in the system; {\em ii}) tag reuse, a measure for the degree to which users reuse a tag perform new annotations; and, {\em iii}) the temporal dynamics of user tag vocabularies, a user-centric analysis of the tag vocabulary evolution over time.

To address the second aspect, we define interest sharing, a metric the activity similarity between a pair of users. Through experiments that compare with a random null model, we show that interest sharing metric captures relevant information about similarity of user preferences, rather than simply coincidence in the tagging activity. Additionally, we present an analysis of the relationship between the implicit ties, as represented by activity similarity between users with respect to their tagging activity, and more explicit ties, such as co-membership in discussion groups and semantic similarity of tag vocabularies.

A summary of the main findings of this study is the following:

\begin{enumerate}

  \item The qualitative characteristics of peer production of information are similar across different systems, but they differ quantitatively, as indicated by the relative levels of item re-tagging and tag reuse.

  \item Interest sharing (the metric that quantifies the similarity between pairs of user's tagging activity) is significantly concentrated on a small fraction of user pairs. This is a characteristic of intelligent choices made by users in tagging systems, and not an implicit result of tagging activity volumes and tag/item popularity distributions as indicated by a comparison of the observed interest sharing distribution to that of a system with the same macro characteristics yet where random tag assignments are used.

  \item User tag vocabularies are constantly growing, but at different rates depending on the age of the user. However, despite the constant increase in size, the relative usage frequency of tags in a vocabulary tends to converge to a stable ranking at early stages of a user's lifetime in the system, as observed in the analysis of the distance between tag vocabularies at different points in time and the user tag vocabulary considering the entire trace.

  \item The implicit and explicit social ties are related, as suggested by the observed higher intensity in group co-membership and tag vocabulary semantic similarity for those user pairs that share interest over items. 
\end{enumerate}

The implications of these results have ramifications along multiple fronts of system design including: {\em i}) recommender systems -- as the concentration of interest sharing on a small fraction of user pairs indicate a highly sparse dataset, it demands more sophisticated techniques to achieve better precision and recall results; {\em ii}) malicious user detection -- as spam detection mechanisms, specially tailored for tagging systems, could use deviations from the characteristics of interest sharing of a non-malicious user population to detect malicious users; {\em iii}) design of distributed infrastructure for tagging systems -- the characteristics of the evolution of user tag vocabularies together with the 'sparse' interest sharing support the intuition that it is possible to design and implement a distributed infrastructure to support tagging features, as it would imply in low communication cost among the parts.

\section{Acknowledgments}

The authors would like thank the Research Computing Center at the University of South Florida for allowing us to use their infrastructure in part of our experiments. Elizeu Santos-Neto was partially supported by the BC Innovation Council Fellowship and AUCC LACREG Exchange Grant. Finally, thanks to Lauro Beltr\~ao Costa and Abdullah Gharaibeh for insightful discussions.

%
%


\bibliographystyle{acmsmall}      
\bibliography{all_2012}   

\end{document}